# Enable High-resolution, Real-time Ensemble Simulation and Data Assimilation of Flood Inundation using Distributed GPU Parallelization

Junyu Wei[1†], Xiangyu Luo[2‡], Weihong Liao[4], Xiaohui Lei[4], Jianshi Zhao[1], Haocheng Huang[3*], Hao Wang[4]

[1] State Key Laboratory of Hydroscience and Engineering, Tsinghua University

[2] School of Civil Engineering and Environmental Science, University of Oklahoma

[3] School of Civil Engineering, Central South University

[4] State Key Laboratory of Simulation and Regulation of Water Cycle in River Basin, China Institute of Water Resources and Hydropower Research

[†] Now at China Devlopment Bank

[‡] Now at Department of Civil and Environmental Engineering, University of Pittsburgh

[*] Corresponding author

# Abstract

Numerical modeling of the intensity and evolution of flood events are affected by multiple sources of uncertainty such as precipitation and land surface conditions. To quantify and curb these uncertainties, an ensemble-based simulation and data assimilation model for pluvial flood inundation is constructed. The shallow water equation is decoupled in the x and y directions, and the inertial form of the Saint-Venant equation is chosen to realize fast computation. The probability distribution of the input and output factors is described using Monte Carlo samples. Subsequently, a particle filter is incorporated to enable the assimilation of hydrological observations and improve prediction accuracy. To achieve high-resolution, real-time ensemble simulation, heterogeneous computing technologies based on CUDA (compute unified device architecture) and a distributed storage multi-GPU (graphics processing unit) system are used. Multiple optimization skills are employed to ensure the parallel efficiency and scalability of the simulation program. Taking an urban area of Fuzhou, China as an example, a model with a 3-m spatial resolution and 4.0 million units is constructed, and 8 Tesla P100 GPUs are used for the parallel calculation of 96 model instances. Under these settings, the ensemble simulation of a 1-hour hydraulic process takes 2.0 minutes, which achieves a 2680× estimated speedup compared with a single-thread run on CPU. The calculation results indicate that the particle filter method effectively constrains simulation uncertainty while providing the confidence intervals of key hydrological elements such as streamflow, submerged area, and submerged water depth. The presented approaches show promising capabilities in handling the uncertainties in flood modeling as well as enhancing prediction efficiency.

# 1 Introduction

Flood inundation poses a great threat to the livelihood of mankind, which makes it an essential issue for hydrology research. Numerical models provide physics-based and continuous analysis for real-world processes and serve as a fundamental tool to predict the emergence and development of flood events. The result of deterministic flood inundation models is typically a series of maps showing the estimated flood extent at different timesteps. However, such results usually contain a certain level of error and fail to represent the probability of different scenarios, which is not optimal for decision-makers. Therefore, efforts need to be made to describe and control the uncertainties in flood modeling, thereby enhancing the reliability of the model predictions.

Researchers have adopted various approaches to deal with the uncertainties in hydrologic and hydraulic models. For example, the method of distributions can be used to derive formulas that describe the evolution of the probability density functions (PDFs) of hydraulic variables (Alawadhi et al., 2018); Bayesian forecasting systems use uncertainty processors to produce probabilistic forecasts based on deterministic models (Han and Coulibaly, 2017). Another common approach is ensemble-based simulation and forecasting, which uses a number of different model realizations to represent the possibilities of various scenarios. The model realizations are essentially Monte Carlo samples, which are drawn from a predefined joint probability distribution of model settings. These samples are then evolved over time and describe the statistical features of model elements. Ensemble-based flood simulation models provide a straightforward representation of uncertainties and have been widely employed in operational forecast systems (Cloke and Pappenberger, 2009), parameter sensitivity analysis (Chaney et al., 2015), and uncertainty quantification (Teng et al., 2017). Besides, the approaches based on polynomial chaos expansion, such as the stochastic Galerkin method (Kusch and Frank, 2018), are also widely adopted for uncertainty quantification. These intrusive methods derive a set of equations for the spatiotemporal propagation of uncertainties, which often (not always) makes them less time-demanding than the Monte Carlo approaches. Nevertheless, additional efforts are also required to rearrange the model codes (Koppel et al., 2017). In this paper, we focus on the efficient implementation of ensemble-based modeling.

To characterize the probability distributions of the model variables, a sufficient number of ensemble members is demanded. Typical ensemble sizes in previous flood prediction studies vary from $10^1$ to $10^2$ (Wu et al., 2020). As the computational complexity of ensemble simulation models grows linearly with the ensemble size, the time efficiency of such models becomes an issue. This issue is especially essential when (a) a real-time prediction of inundation is required to maximize the time for evacuating personnel and assets; (b) the model is built with a high resolution to simulate the processes at fine scales, which further increases the computational complexity exponentially. In such cases (for example, urban pluvial inundation modeling), a modeler must utilize more computational resources using parallel computing techniques to accelerate the simulation process.

Naive implementations of numerical models usually use one processor only. Parallel computing divides the computation task into smaller ones, which are assigned to multiple processors to fully exploit their capabilities. Conventional parallel techniques using central processing unit (CPU), such as message passing interface (MPI) which is based on processes, and open multi-processing (OpenMP) which is based on threads, have been widely adopted in two-dimensional (2-D) hydraulic models. Sanders et al., (2010) used the single-process-multiple-data parallel paradigm in a shallow water model and achieved satisfying scaling performance in 512 processors. Noh et al., (2016) studied the impact of parameter uncertainty in urban flood modeling with an ensemble of 2,000 members and accelerate the computation using MPI. For large-scale, high-resolution flood simulation problems, a valid approach is to decompose the studied district into multiple subdomains and assign the computation of different domains to different processes using MPI (Sanders et al., 2019). By accelerating each MPI task with OpenMP, a speedup of 153 was reached in the flood inundation simulation of a 35 km × 37 km area using 8,192 CPUs (Kobayashi et al., 2015). A similar method was also adopted in hydrological modeling, achieving speed-up ratios ranging from 2 to 70 in river basins of different sizes (Vivoni et al., 2011). Moreover, the hybrid MPI+OpenMP method has also been proven useful in three-dimensional computational fluid dynamics models (Afzal et al., 2017) at various scales and cases.

Apart from CPU-based parallelization, the last decade has also witnessed a surge of applications of GPU in scientific simulation models. GPU is firstly designed to deal with matrix-shaped data that correspond to pixels on the computer screen. Featuring good performance in computation-intensive and data-intensive tasks, GPU has gained popularity in the scientific computing community, especially in hydrologic and hydrodynamic models (Smith and Liang, 2013; Lacasta et al., 2015; Aureli et al., 2020). A comparison of the performance of different devices running numerical weather models showed that GPU-based programs are 1.4-2.5 times faster than those based on CPU (Govett et al., 2017). CUDA, which is proposed by Nvidia for heterogeneous computing and utilizes both CPU and GPU hardware, is a convenient and powerful tool for modelers to transform their CPU programs to GPU ones with limited changes to the source codes. Kalyanapu et al. (2011) developed a simulation software Flood2D-GPU using CUDA and reached a speedup of 88 in a dam break problem with 708k computation cells. Vacondio et al., (2014) used GPU to accelerate the computation of a finite-volume-based 2-D shallow water equation solver and achieved a 200× speedup. For large-scale cases, the computational load and data amount might exceed the capacity of a single GPU, and the multi-GPU parallelization paradigm must be adopted. The computation domain is divided into a number of sub-domains (as stated above) and the computation and variables of each domain are assigned to the corresponding GPU. A catchment-scale fluvial flood modeling system based on multi-GPU parallel computing was proposed by Xia et al. (2019). By utilizing 8 Nvidia Tesla K80 GPUs, the real-time simulation of a 100-million-cell hydrodynamic model at 5-meter resolution was achieved. Further, the GPU-accelerated full hydrodynamic model is integrated with numerical weather predictions to construct a highly efficient flood forecast system based on catchment-scale modeling (Ming et al., 2020). For cases with an even larger scale, the modeler must make use of multiple physical nodes in supercomputing centers, and the MPI technique is necessary for communication between the nodes. Sharif et al. (2020) presented a state-of-the-art evaluation of the performance of 2-D flood models on heterogeneous high-performance-computing architectures. Both finite-volume and finite-difference schemes were tested, and up to 768 GPUs were employed. The simulation of a 5-day flooding event with 272 million cells and a 5-meter resolution was completed in 50 minutes. Besides, the open-source TRITON model presented by Morales-Hernandez et al., (2021) supports both CPU and GPU clusters to realize high-efficiency hydraulic computations at large spatial and temporal scales. These studies mainly focus on implementing the GPU parallelization of a deterministic flood simulation model. The application of GPU in ensemble-based inundation simulation is relatively rare.

Apart from the efficiency issue, the accuracy of predictions is another essential aspect of flood modeling. For a specific model structure, the error in model predictions stems from the input factors such as rainfall, initial and boundary conditions, and the physical characteristics (e.g., roughness and infiltration capacity) of the underlying surface. By sampling these factors from a prior guess of their probability distributions, an ensemble of model realizations can be built to represent various sources of uncertainties. Nevertheless, it is difficult to cover all possible circumstances in the initial ensemble, and the ensemble results might deviate from the real state. Data assimilation (DA) is the procedure of fusing observations of system states (e.g., hydrological variables such as water depth and flow rate) with model predictions and generating the best estimation of the model state fields. In previous literature, DA approaches have served as effective tools to reduce model error and enhance output reliability in ensemble forecasts (Anderson and Collins, 2007; Nerger and Hiller, 2013; Liu et al., 2020).

Variants of the Kalman filter and particle filter (PF) are the most widely adopted ensemble DA methods. For example, the ensemble Kalman filter (EnKF) (Evensen, 2009), the ensemble-square root Kalman filter (Whitaker and Hamill, 2002), and the singular evolutive interpolated Kalman filter (Pham, 2001) have been proven effective in geophysical DA systems. In flood simulation, the EnKF is also a common approach for the estimation of model states and parameters (Chen et al., 2013; Barthélémy et al., 2017; Cooper et al., 2019; Ziliani et al., 2019). Despite its popularity, the reliability of EnKF in strongly nonlinear cases may be inferior due to its linear estimation scheme. The PF, which directly originated from the Bayesian theory (Arulampalam et al., 2002), offers an alternative with better adaptation in nonlinear systems. It has been employed successfully in 1- and 2-D hydrodynamic models (Xu et al., 2017; Cao et al., 2019) as well as hydrological models (Liu et al., 2012). For instance, Abbaszadeh et al. (2020) used a modified PF to assimilate both soil moisture and streamflow observations to improve the predictions of a coupled weather-hydrological model, and Zarekarizi (2018) investigated the performance of PF in predicting the hydrological variables in several land surface models.

The application of parallelization in DA is necessary when the numerical model has a high computational complexity, especially for ensemble DA which features good parallelism inherently. Nerger and Hiller (2013) studied the parallelization strategies and their scalabilities of the Kalman-filter-based ensemble DA algorithms. MPI was frequently used to run multiple model realizations in parallel (Nerger et al., 2005; Kurtz et al., 2017) and accelerate the evolution of the ensemble members. On the model level, some previous studies of ensemble DA (Hostache et al., 2018; Zarekarizi, 2018; Liu et al., 2020) use techniques such as OpenMP and CUDA to accelerate the integration of the physical process. Despite this, little attention has been paid to how GPU-based distributed parallelization could facilitate the ensemble DA in pluvial inundation models.

In this paper, we manage to take the advantage of state-of-the-art parallelization techniques to construct a real-time probabilistic pluvial inundation model. The 2-D high-resolution hydrodynamic model is built using an efficient discretization scheme that is suitable for GPU parallelism. An ensemble of model realizations is initialized with different input factors, and the parallelization was carried out in two hierarchies at both model cell and ensemble member levels. Synthetic observations of hydrological variables are assimilated using Bayesian update rules to reduce model biases and control uncertainties. In section 2, the numerical method and the DA algorithm are introduced. In section 3, we explain the parallel implementation details of the simulation and assimilation models. In section 4, the model is tested with real-world topographic data and the simulation results are presented. Conclusions are made in section 5.

## 2 Numerical methods

### 2.1 Inundation model

The flood inundation process is commonly simulated using models based on the 2-D shallow water equations (SWE). The full solutions of the SWE, including finite difference and finite volume schemes (Sharif et al., 2020), make comprehensive descriptions of the physical process but are also quite time-demanding. A simplified and efficient way of solving flood inundation problems is the storage cell method (Bates and De Roo, 2000). This method discretizes the studied area into computation cells and solves the continuity equation of volume in each cell:

$$\frac{\Delta h}{\Delta t} = \frac{\Delta V}{\Delta x \Delta y} \tag{1}$$

where $h$ [L] and $V$ [L$^3$] indicates the water depth and volume of a cell, respectively. In this paper, we consider a cartesian partition of the cells, thus $\Delta x$ [L] and $\Delta y$ [L] are the size of the cell at the $x$ and $y$ directions, respectively. $\Delta t$ [T] is the time step of the simulation. The change of volume $\Delta V$ is determined by the flux exchange with the four adjacent cells, and the flux between two cells can be calculated using the inertial form of the Saint-Venant momentum equation (Sridharan et al., 2020):

$$q_{t+\Delta t} = \frac{q_t - gh_{\text{flow}}\Delta t S_{\text{surf}}}{\left(1 + gn^2|q_t|\Delta t/h_{\text{flow}}^{7/3}\right)} \tag{2}$$

where $q_t$ [L$^2$T$^{-1}$] is the flow per unit width on the cell boundary at time $t$, $S_{\text{surf}}$ [1] is the slope of the water surface, $g$ [LT$^{-2}$] and $n$ [L$^{-1/3}$T] indicates gravity and friction coefficients, respectively; $h_{\text{flow}}$ is the current flow depth. The flow depth indicates the depth of the water flow between the cells, which is defined as:

$$h_{\text{flow}} = \max(h_1 + z_1, h_2 + z_2) - \max(z_1, z_2) \tag{3}$$

where $h_1$, $h_2$ and $z_1$, $z_2$ indicate the water depths and bed elevations of the two cells, respectively. The setting of timestep is subject to the Courant-Freidrichs-Levy condition and is given by:

$$\Delta t_{\max} = \frac{k \Delta x}{\sqrt{g h_{\max}}} \tag{4}$$

where $\Delta t_{\max}$ is the maximum timestep, $h_{\max}$ is the maximum water depth in the simulation region, and the coefficient $k$ usually lies in the range of 0.2~0.7 (Bates et al., 2010). After calculating the fluxes between cells, the volume change can be derived:

$$\Delta V_{(r,c)} = Q_{x,(r,c)} - Q_{x,(r+1,c)} + Q_{y,(r,c)} - Q_{y,(r,c+1)} \tag{5}$$

where $(r, c)$ indicates the indices of row and column, $Q_x$ and $Q_y$ represents the fluxes in the $x$ and $y$ directions, respectively. In this way, a local inertial form of storage cell inundation model is built. This numerical scheme features looser constraints on the timestep compared with diffusive schemes and a simpler calculation process compared with the full solution, so it is widely adopted in well-known inundation models such as LISFLOOD-FP and flood prediction studies (Dottori and Todini, 2011; Shustikova et al., 2019; Sridharan et al., 2020). Notably, Gozzolinio et al. (2019) argued that the inertial scheme has certain physical limitations on the wet-dry front. Nevertheless, previous studies have shown that it produces reliable results in both experimental and real-world applications (Bates et al., 2010; De Almeida and Bates, 2013; Sridharan et al., 2020) at spatial resolutions of up to 2 meters, making it useful in fast-warning modeling systems.

## 2.2 DA scheme

In this paper, we use the PF to sequentially assimilate hydrological observations and control model uncertainties, because it is less affected by nonlinearity than Kalman filter-based methods. Generally, we use an ensemble of simulation models to represent the uncertainties introduced by the model inputs. The ensemble of models evolves as:

$$\boldsymbol{x}_{i,t+\Delta t} = \boldsymbol{f}(\boldsymbol{x}_{i,t}, \boldsymbol{u}_{i,t}, \boldsymbol{p}_{i,t}) \qquad i = 1, \dots, N \tag{6}$$

where $\boldsymbol{x}$, $\boldsymbol{p}$, and $\boldsymbol{u}$ represent model variables, parameters, and external forces, respectively. $\boldsymbol{f}$ is the model operator. The subscript $i$ indicates the ensemble member (particle) index, and $N$ is the ensemble size.

At time 0, the model is initialized by extracting $\boldsymbol{x}_{i,0}$ and $\boldsymbol{p}_{i,0}$ from predefined prior distributions, while the input series $\{\boldsymbol{u}_{i,t}\}$ is also initialized. The PDF of the model state $p(\boldsymbol{x}, \boldsymbol{u}, \boldsymbol{p})$ is described by these Monte-Carlo samples:

$$p(\boldsymbol{x}, \boldsymbol{u}, \boldsymbol{p}) \approx \sum_{i=1}^{N} w_i \delta\big((\boldsymbol{x}, \boldsymbol{u}, \boldsymbol{p}) - (\boldsymbol{x}_i, \boldsymbol{u}_i, \boldsymbol{p}_i)\big) \tag{7}$$

where $w_i$ is the weight of particle $i$ and $\delta(\cdot)$ is the Dirac delta function. Initially, the weights of all particles are equal. The models then evolve independently until the observation comes in when the model state PDF is updated following the Bayes rule:

$$p^+(\boldsymbol{x}, \boldsymbol{u}, \boldsymbol{p}|\boldsymbol{d}) \propto p(\boldsymbol{d}|\boldsymbol{x}, \boldsymbol{u}, \boldsymbol{p}) p^-(\boldsymbol{x}, \boldsymbol{u}, \boldsymbol{p}) \tag{8}$$

where $p^+(\boldsymbol{x}, \boldsymbol{u}, \boldsymbol{p})$ and $p^-(\boldsymbol{x}, \boldsymbol{u}, \boldsymbol{p})$ are the posterior and prior PDFs respectively and $p(\boldsymbol{d}|\boldsymbol{x}, \boldsymbol{u}, \boldsymbol{p})$ is the likelihood of the observation $\boldsymbol{d}$ conditioned by the model state. One can map the model variables $\boldsymbol{x}$ to a corresponding vector $\boldsymbol{y}$ in the observation space (e.g., calculate the water level at gauge stations based on simulated water depth distribution), and the likelihood can be defined (e.g., using a Gaussian distribution) as:

$$p(\boldsymbol{d}|\boldsymbol{x},\boldsymbol{u},\boldsymbol{p}) = p(\boldsymbol{d}|\boldsymbol{x},\boldsymbol{y},\boldsymbol{u},\boldsymbol{p}) = \frac{\exp\left(-\frac{1}{2}(\boldsymbol{y}-\boldsymbol{d})^\mathrm{T}\boldsymbol{\Sigma}^{-1}(\boldsymbol{y}-\boldsymbol{d})\right)}{\sqrt{(2\pi)^{N_d}|\boldsymbol{\Sigma}|}} \qquad (9)$$

where $N_d$ is the number of observations, and $\boldsymbol{\Sigma}$ is the error covariance matrix of $\boldsymbol{d}$. Usually, the errors are independent and $\boldsymbol{\Sigma}$ is diagonal. In this way, the posterior weights of the particles can be evaluated:

$$\tilde{w}_i^+ = p(\boldsymbol{d}|\boldsymbol{x},\boldsymbol{u},\boldsymbol{p})w_i^- \qquad (10)$$

$$w_i^+ = \frac{\tilde{w}_i^+}{\sum_i \tilde{w}_i^+} \qquad (11)$$

where $\tilde{w}_i^+$ and $w_i^+$ are the raw and normalized particle weights, respectively. This formulation of PF sometimes suffers from particle degeneracy, which means the weights of most particles become close to zero and the ensemble loses its diversity. Specifically, the diversity of updated particles can be evaluated by the number of effective samples $N_\mathrm{eff}$:

$$N_\mathrm{eff} = \frac{1}{\sum_i w_i^{+2}} \qquad (12)$$

If $N_\mathrm{eff}$ is very small, particle degeneracy can not be ignored. To alleviate this problem, the systematic resampling technique (Arulampalam et al., 2002) is employed by duplicating and perturbing high-weight particles.

# 3 Parallel implementations

To enable more scientific calculation programs with GPU parallelization, CUDA supports both C/C++ and Fortran languages. In this paper, we use CUDA Fortran to build the high-performance simulation model.

## 3.1 Deterministic model

The ensemble-based simulation model consists of multiple independent deterministic models, which are described here. The presented model evolves the following steps in one numerical iteration: (I) Update and assign precipitation inputs to all computation cells, (II) Update the boundary conditions, (III) Calculate the fluxes between adjacent cells, and (IV) Calculate the volume and water depth of all computation cells, of which steps (I) and (II) are conducted every 5 minutes. For a serial implementation on CPU, all calculations are executed sequentially by a single processor core. On the other hand, for a GPU-based parallel implementation, the simulation data must be transferred to the GPU's independent memory space. The steps (I) ~ (IV) are executed by four corresponding CUDA kernels and the computation of different cells is assigned to massive threads and run by numerous multiprocessors. In CUDA, the computation threads are organized in a three-layer nested form of 'grid' - 'block' - 'thread', which is shown in Fig. 1. The indices of blocks and threads can be one-, two-, or three-dimensional, which easily correspond to the dimensions in the real-world space. Therefore, the computational cells in the 2-D hydrodynamic model are indexed using 2-D cartesian coordinates, forming a topological structure shown in Fig. 2. It can be seen that the links which connect adjacent cells are classified into two groups $Lx$ and $Ly$ according to their directions, and both groups of links are also indexed by 2-D cartesian coordinates. Hence, the data of cells (e.g., cell elevation) and links (e.g., link roughness) are stored in 2-D arrays. With such definitions of topology and data structure, we assign the computation of cell $C(r,c)$ to thread $T(r,c)$ (assume there is only one block), so that the indices for cells, links, data arrays, and threads are all aligned. Specifically, the calculation of the volume change in $C(r,c)$ depends on the fluxes on links $Lx(r,c)$, $Lx(r,c-1)$, $Ly(r,c)$, and $Ly(r-1,c)$, which require the water depths in $C(r,c)$ and $C(r,c+1)$, $C(r,c-1)$, $C(r+1,c)$, and $C(r-1,c)$, respectively. In this way, the neighboring threads are occupied with the calculation of neighboring cells and links and access neighboring data, which matches the technical characteristic of CUDA and benefits computation efficiency.

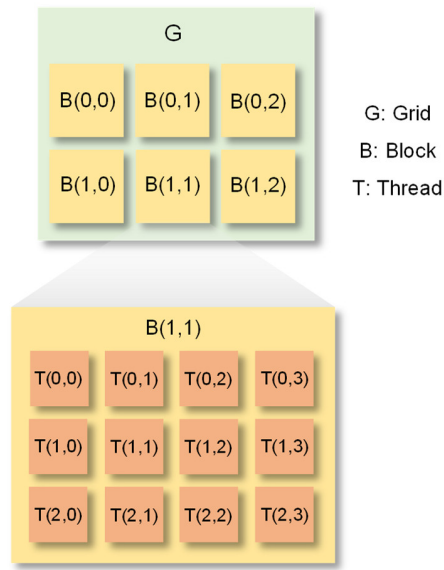

**Figure 1.** An example of the parallel architecture of the CUDA programming model. One grid corresponds to a single GPU device. The 2-tuples in the bracket indicate the indices of the blocks or threads. Here the grid consists of 6 blocks (2 rows × 3 columns) and each block contains 12 threads (3 rows × 4 columns).

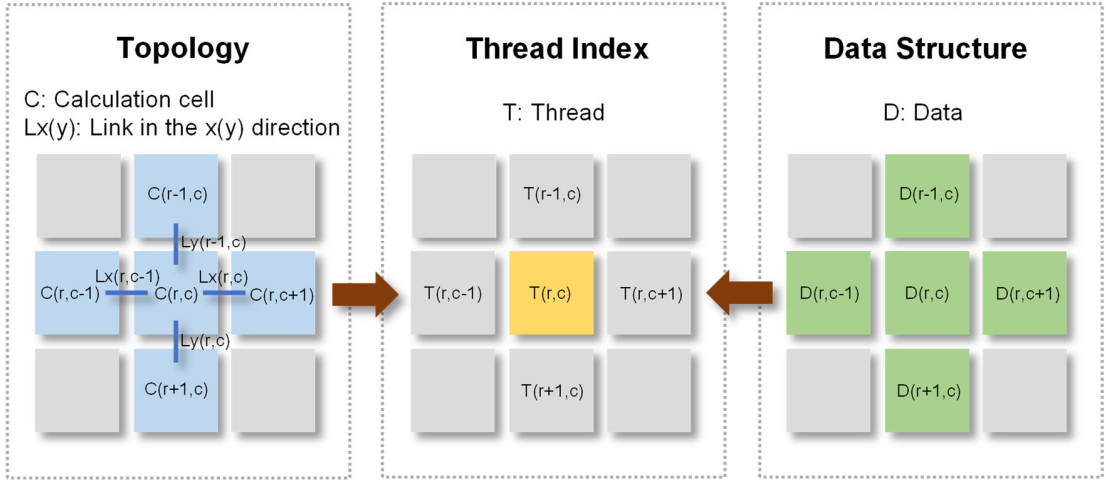

**Figure 2.** Parallel design of the deterministic simulation model.

**3.2 Ensemble model**

Creating an ensemble of parallel model realizations increases the computation complexity by $10^1 \sim 10^2$ times, depending on the ensemble size $N$. If all ensemble members are completely independent, both the time and space complexities of the ensemble model grow linearly with $N$. Nevertheless, there are ways through which we can reduce these complexities.

When building a probabilistic model, a modeler must decide which model parameters are considered uncertain and which are considered certain. The uncertain parameters will be sampled in the parallel realizations, while the definitive ones will not. In the deterministic model, all of the space-distributed parameters (whether uncertain or not) and model state variables are stored in 2-D arrays. In the ensemble simulation model, the definitive parameters can be shared by multiple model realizations, which avoids unnecessary memory usage. Meanwhile, the uncertain parameters and variables differ for each Monte-Carlo sample, so an additional data dimension is required. We extend these arrays as well as the thread indices to 3-D, shaped as $R_{\text{en}}(i, r, c)$ and $T(i, r, c)$ respectively. Specifically, the $R_{\text{en}}$ arrays in our probabilistic prediction model include the water depth, precipitation, and drainage capacity of computation cells, as well as the fluxes and Manning's coefficients of cell links. It is notable that arrays in CUDA Fortran are stored in column-major order, so that the data of different ensemble members with identical coordinates have continuous memory addresses, as shown in Fig. 3. This implementation ensures that the indices of threads and data arrays are kept aligned so that neighboring threads access neighboring data. Besides, compared with the deterministic model, the additional memory accesses of the ensemble-based simulation program are continuously 'stretched' from the original addresses. Such a pattern is beneficial for improving the cache hit ratio and enhancing simulation efficiency.

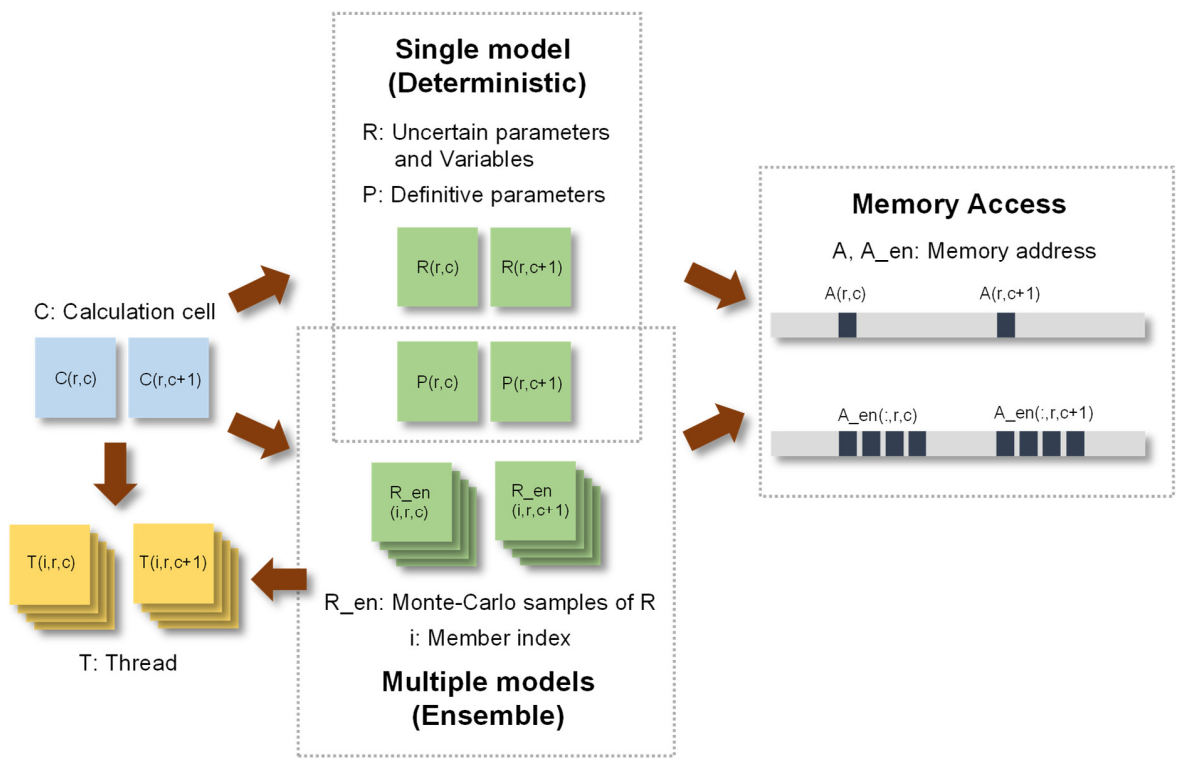

**Figure 3.** The differences between ensemble and deterministic simulation models.

**3.3 Heterogeneous computing design**

In CUDA's definitions of heterogeneous computing, CPU and GPU are referred to as host and device respectively. An essential concern in deploying the models on a distributed-memory system is reducing the communication overhead, i.e., the latency caused by transferring data between host and devices. In deterministic simulations, the host is responsible for preprocessing, after which the simulation data is uploaded to the device (Fig. 4). This ensures that no further communications are required between host and device in the model iteration.

To fully utilize multiple GPUs in the ensemble simulation, how the model realizations are distributed among the devices must be designed carefully. Here we follow the natural approach of decoupling the ensemble, i.e., dividing the ensemble into $m$ groups ($m$ equals the number of devices employed) and assigning the data and calculation of each group to one device. The initial ensemble is generated and grouped on the host. Subsequently, the model groups are sent to their corresponding devices (Fig. 4), and no data communication is required for the simulation process.

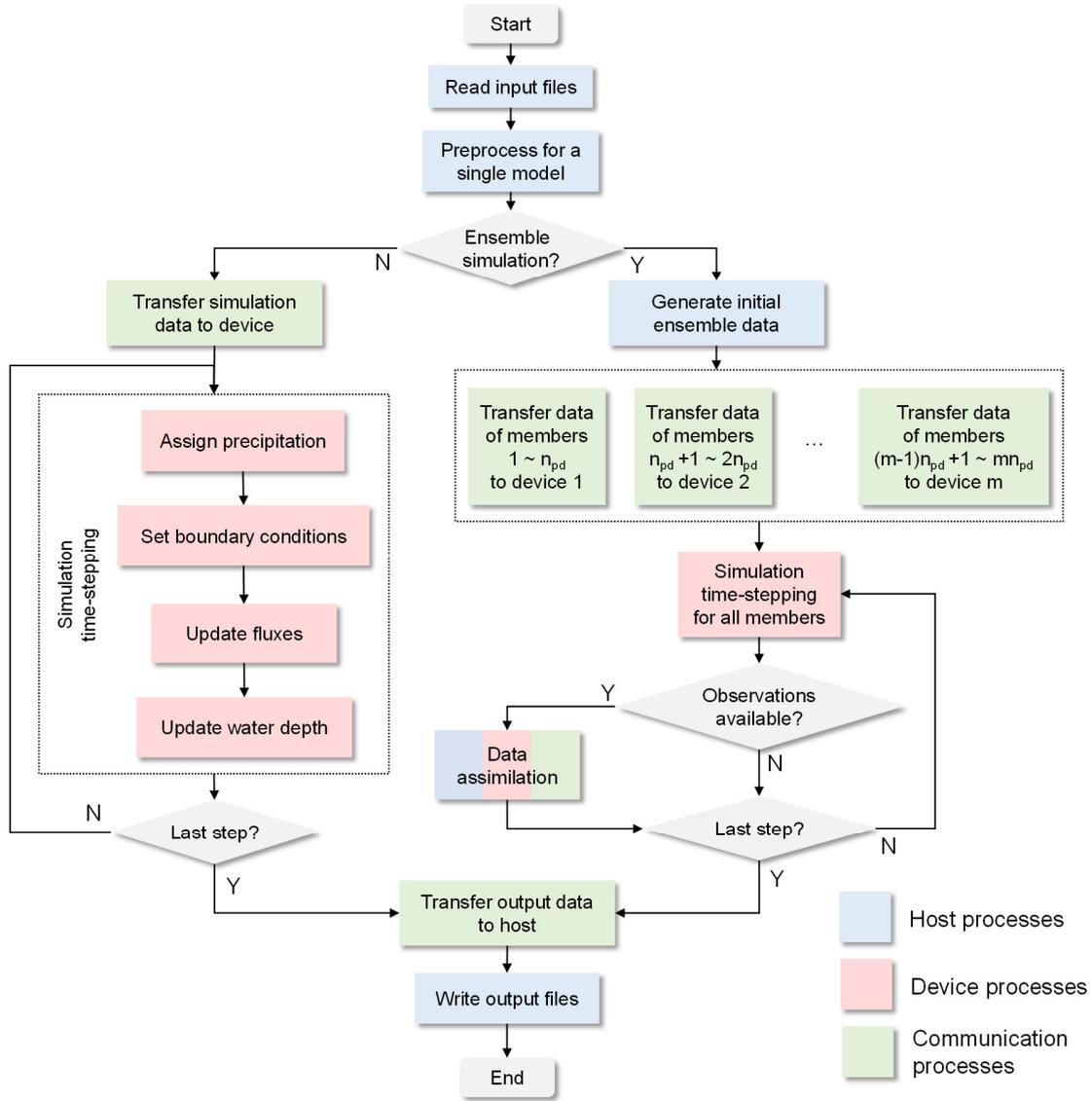

**Figure 4.** Flowchart of the entire simulation and DA process. $n_{\text{pd}}$ indicates the number of model realizations on each device. The detailed steps of DA are further illustrated in Fig. 5.

On the other hand, the calculations in the assimilation process involve five steps: (I) For each ensemble member $i$, generate model outputs $\boldsymbol{y}_i$ based on the simulated state field, (II) Calculate the member weights, (III) Evaluate $N_{\text{eff}}$ using Eq. (12), and execute the next two steps if $N_{\text{eff}}$ is smaller than a given threshold $N_{\text{thr}}$, (IV) Resample, i.e., decide the number of duplicates of each particle in the posterior ensemble; (V) Replace ensemble realizations with the resampled ones. It can be easily seen that steps (II) and (V) require non-local data in a distributed implementation of the ensemble model. In this paper, we use the host as the pivot of gathering and distributing data associated with assimilation. Specifically, the ensemble model outputs $\boldsymbol{y}_i$ are collected on the device and transferred to the host in step (I), and steps (II) and (III) are executed by the CPU. Subsequently, if $N_{\text{eff}} < N_{\text{thr}}$, step (IV) is executed on the host and the $R_{\text{en}}$ arrays on all devices are transmitted to the host. Next, for step (V), the host arrays $R_{\text{en}}(i,:,:)$ are copied for $k$ times ($k$ is the number of duplicates of member $i$) and transmitted back to the devices, replacing the $R_{\text{en}}$ arrays of $k$ ensemble members. These steps are illustrated in Fig. 5. The data transfer is optimized by executing the memory copies between different devices and the host asynchronously, e.g., the memory copy $\text{Device1} \to \text{Host}$ is initiated before the memory copy $\text{Device0} \to \text{Host}$ is completed. In addition, the value of $N_{\text{thr}}$ ranges from $N/3$ to $N$ according to previous studies (Doucet et al., 2000; Moradkhani and Hsu, 2005). A large value of $N_{\text{thr}}$ allows frequent resampling so that particle degeneracy can be better addressed, but it is also time-consuming because steps (IV) and (V) require massive data transfer. Here we set $N_{\text{thr}} = 0.8N$ to balance the resampling performance and calculation burden.

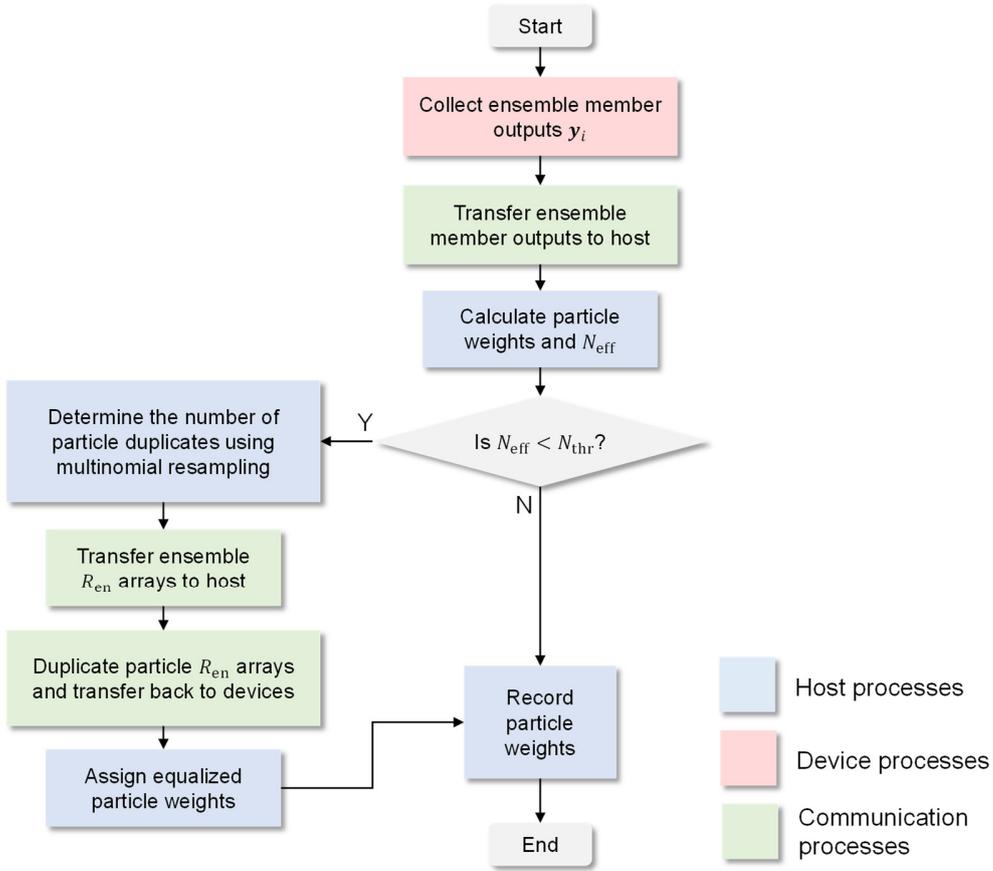

**Figure 5.** Flowchart of a single round of DA calculations.

## 4 Simulation and DA results

### 4.1 Test case

We choose an urban area in the city of Fuzhou, China to investigate the performance of the ensemble simulation and DA model. Predicting city pluvial inundation demands both high-resolution results and time efficiency, therefore is a suitable case to test the model. The district map, as well as the elevation and land usage status, are illustrated in Fig. 6. The digital elevation map (DEM) and land usage data provided by the Fuzhou Investigation and Surveying Institute are both raster-based with a resolution of 3 meters. The computation cells are constructed directly upon these data, forming a cartesian grid of 3-meter spatial resolution. The entire map has an area of 8.2 km × 7.6 km. The cells outside the simulation region are not considered in the simulation, and the total number of simulated cells is 4,002,457. In the current version of the model, we set a fixed time step of 0.25 s for inundation calculation. As the maximum simulated water depth in the following cases is 4.2 m, this timestep does not exceed the maximum timestep defined in Eq. (4) given $k = 0.54$. The water body cells are initialized with an equal water level of 3.0 meters. The boundary cells at the two stream outlets (near the streamflow gauges 1 and 2 in Fig. 6) are assigned with synthetic hydrographs (for the lack of real-world measurements). These boundary hydrographs correspond to the flooding process and initial condition and can be fetched via the data link in the acknowledgments. Other boundary links are assigned with a zero-flux Neumann-type boundary condition, which is a common simplification in city inundation studies (Chen et al., 2009; Wang et al., 2018). For all simulations, we set a 2-hour warmup period in which the model runs with no rainfall.

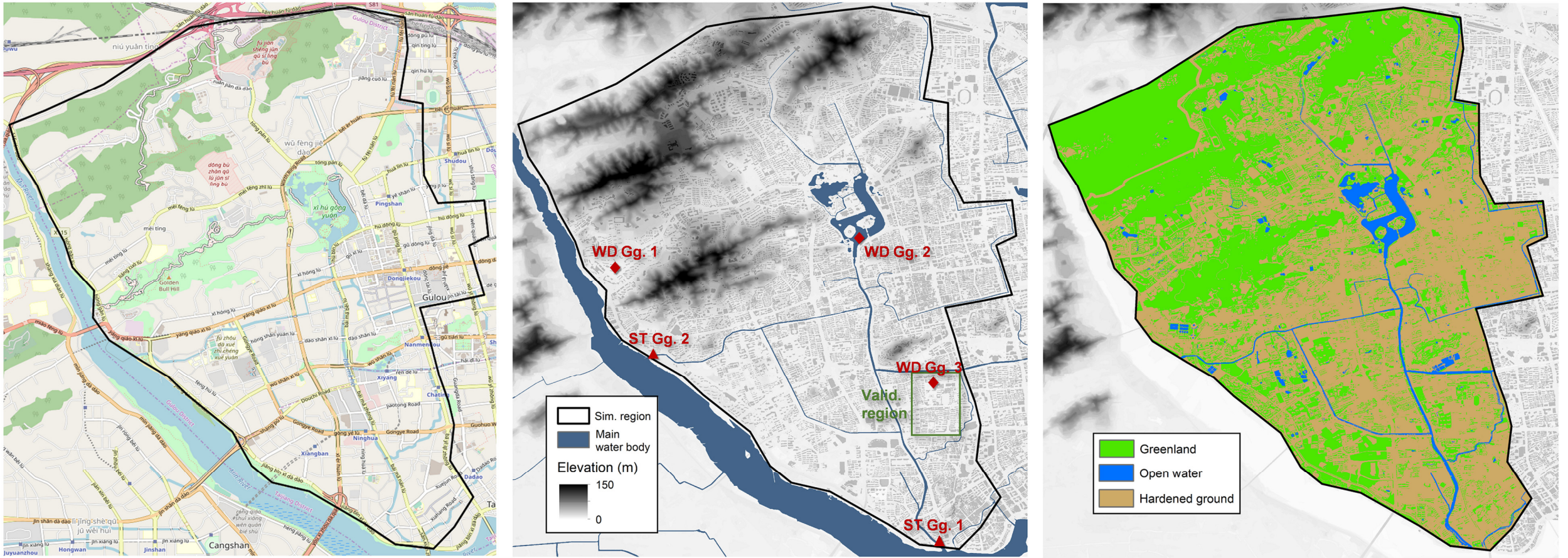

**Figure 6.** The study area. The left figure shows the district map (captured from OpenStreetMap). The middle figure shows the DEM data. The right figure is the simplified land cover map. In the middle map, stream gauges (ST Gg.s) and water depth gauges (WD Gg.s) are shown by triangles and diamonds respectively. The study region for model validation is marked by the green rectangle.

The major external driving force in this pluvial inundation scenario is precipitation. We choose a typical storm rainfall process of Fuzhou (Fig. 7) as the input of the model. Different land cover types in Fig. 6 have different physical properties. Specifically, the roughness coefficients of hardened ground, green land, and water bodies are illustrated in Table 1. The green land area has a runoff coefficient of 0.9, which means 10% of the precipitation is lost in infiltration. The underground pipe network in urban areas has a certain drainage capacity, which is represented in the model by assigning a fixed infiltration rate $I_{\text{drn}}$ to the hardened ground district. $I_{\text{drn}}$ is set to 12 mm/h which is estimated based on local inundation data. No infiltration is considered in water bodies. The friction coefficients of three land cover types, the drainage capacity $I_{\text{drn}}$ and the precipitation series are considered uncertain input factors in ensemble simulation. Random perturbations and white noises are added to these inputs, creating the Monte Carlo samples of the initial ensemble.

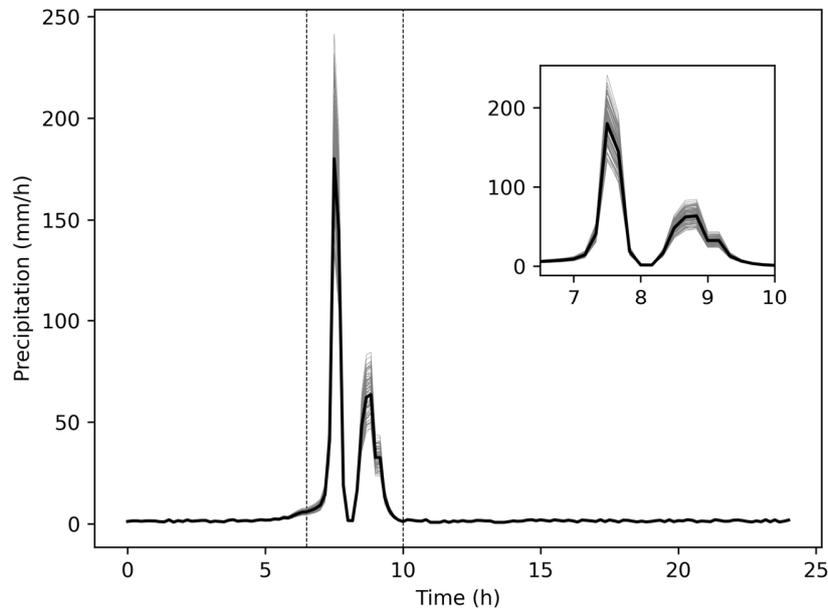

**Figure 7.** Precipitation time series used for simulations. The inner plot captures a segment from 6.5 h to 10 h from the complete series. The black line is the typical storm rainfall process used in the deterministic

model. The gray lines are samples perturbed from the black line using a Gaussian distribution and are employed as the inputs of ensemble simulation.

**Table 1.** Roughness coefficients of different land cover types (unit: $sm^{-1/3}$). The values for the deterministic model are set to similar values as in previous studies (e.g., Gironás et al., 2010).

|  | Green land | Water bodies | Hardened ground |
| --- | --- | --- | --- |
| Deterministic | 0.15 | 0.020 | 0.017 |
| Initial ensemble | 0.08 ~ 0.22 | 0.016 ~ 0.030 | 0.010 ~ 0.022 |

## 4.2 Model validation

Firstly, we use the well-developed Infoworks ICM model (HR Wallingford) to validate the presented model. Infoworks ICM supports an integrated 1-D and 2-D hydrodynamic modeling framework, and is widely adopted in urban inundation modeling (Fan et al., 2017; Zhou et al., 2019). Here we only use the 2-D surface model in ICM, which solves the shallow water equation with a finite-volume numerical scheme according to its manual. The test area for validation is located southeast of the entire study region (Fig. 6), and its elevation map is demonstrated in Fig. 8 (a). The area is divided into 67,540 computational cells using a 3-meter spatial resolution. A 24-hour typical rainfall event (Fig. 7, black line) and a dry initial condition are applied to this district, forming a 2-dimensional inundation process. The final water depth field simulated by the presented model and ICM are demonstrated in Fig. 8 (b) and Fig. 8 (c) respectively. The results show that the inundation maps predicted by the two models are very similar, with only slight differences in several small areas. Fig. 8 (d) shows the absolute water depth difference is within several centimeters in most locations, with a root mean squared difference of 3.2 cm.

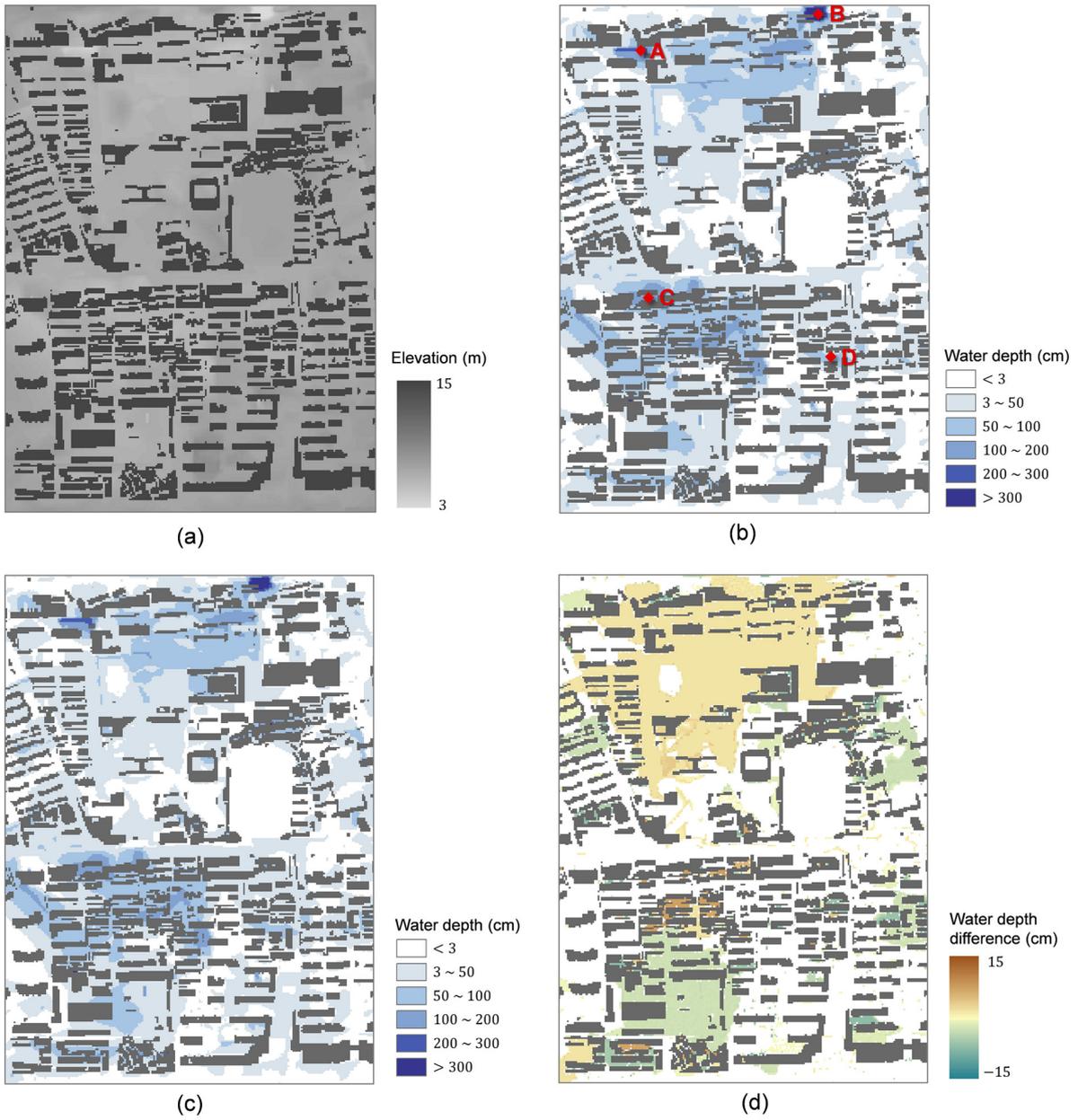

**Figure 8.** Validation of the presented model. Subplot (a) is the test region DEM. Subplots (b) and (c) show the final inundation map predicted by the presented model and Infoworks ICM respectively, while subplot (d) shows the difference between the two models.

The hydrographs at four heavily inundated locations (Fig. 8b) are compared in Fig. 9, which also shows a good coincidence between the two models. These results indicate that the presented model is capable of making reliable predictions in urban inundation events. In addition, we also check the output's sensitivity to the precision of computation. Single and double precisions are used to run this case respectively, and no significant differences are found. Specifically, the maximum absolute water depth difference between the two runs is 0.3 cm and the root mean squared difference is 0.04 cm.

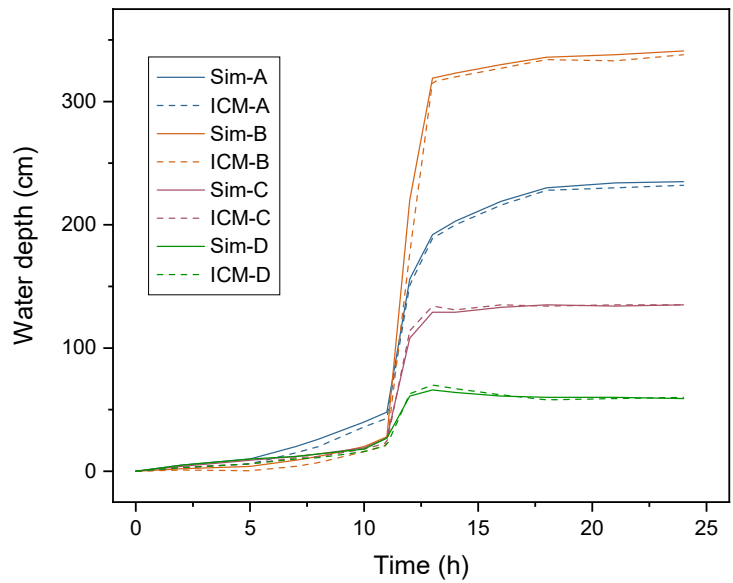

**Figure 9.** Comparison of the hydrographs predicted by the presented model (solid lines) and Infoworks ICM (dashed line). The investigated points A, B, C, and D are shown in Fig. 8 (b).

### 4.3 Deterministic model

In this study, we use an Nvidia Tesla P100 GPU (16GB memory, 3,584 CUDA cores, each with a base frequency of 1.33 GHz) to accelerate the deterministic model, while serial codes based on the same mathematical formulations are executed by an Intel Xeon E5-2650 v4 CPU (12 cores, each with a base frequency of 2.20GHz). Note that although we use a multi-core CPU, only one core is used by the serial codes in comparison. Since few distinctions are found between the single-precision and double-precision runs in model validation, all programs below use single precision to save computational burden. The dimension of thread block $dim_B = (dim_x, dim_y, dim_z)$ is set to (96,1,1) to optimize memory access. The run time of simulating a 24-hour inundation process is 5.9 min and 1317.6 min for the parallel (GPU) and serial (CPU one-core) codes respectively, resulting in a $223 \times$ speedup.

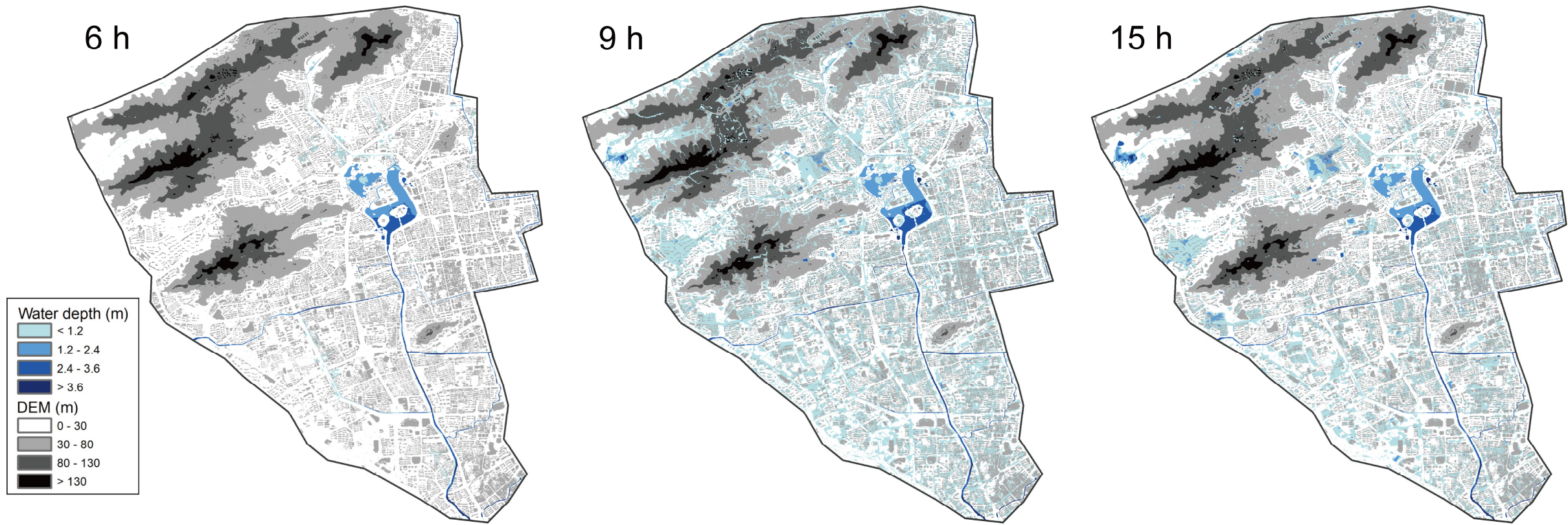

**Figure 10.** Water depth distributions at times 6 h, 9 h, and 15 h simulated by the deterministic forecast model. Cells with water depth < 10 cm are not shown.

The spatial distribution of water depth at hours 6, 9, and 15 is shown in Fig. 10. It can be seen that before the main peak of precipitation arrives, only very few districts are slightly inundated. At hour 9, inundation areas are distributed across the whole simulation region, and runoff convergence can be observed clearly in northern mountainous districts. The confluence of surface flows nearly ends at hour 15, resulting in many regional flooded areas which are slowly drained by the pipe network.

**4.4 Ensemble-based simulation and DA**

The hardware employed for ensemble forecasting includes the aforementioned CPU and $8\times$ Tesla P100 GPUs. In Monte Carlo simulations, an ensemble size of up to $10^4$ is needed to construct the PDF very precisely (Alawadhi et al., 2018). Nevertheless, considering the computation burden and the relation with ensemble weather forecasts, the ensemble sizes adopted by most of the ensemble-based flood models are smaller than 100 (Wu et al., 2020). In this study, we created an ensemble of 96 concurrent model realizations (12 models on each GPU) with the same spatial and temporal resolutions as the deterministic model and varying model inputs as stated in 4.1. Therefore, the total number of computational cells that are updated simultaneously is 384 million. As the dimension of data arrays increases, the third dimension of thread blocks $dim_z$ is set to the number of model realizations on each device. Different values of $dim_x$ are tested to obtain the best performance, and the optimized block dimension is set to $dim_B = (8,1,12)$. Under these settings, the probabilistic prediction of a one-hour inundation process takes 118 s on average, which is more than 30 times faster than real time. The serial version using one CPU core and single precision is not tested because it is too time-consuming.

Nevertheless, if we assume that the serial time consumption grows linearly with the ensemble size, the speedup achieved by multi-GPU parallelization in this case is 2680×.

We investigate the scaling effect in terms of both ensemble members and devices. Firstly, device scalability is tested by measuring the run time when different numbers of devices are employed. Here, each device is assigned 12 model realizations simulating a 1-hour inundation process. It is demonstrated in Fig. 11 that increasing the number of devices brings little additional overhead and the ensemble simulation model has good device scalability.

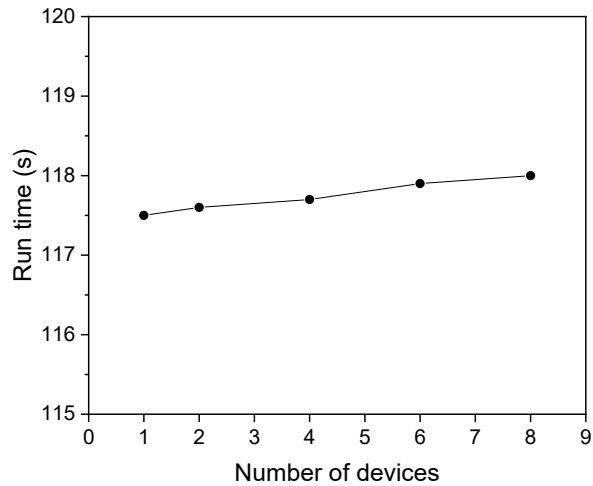

**Figure 11.** Device scalability of the ensemble simulation model ($n_{\mathrm{pd}} = 12$).

On the other hand, when the number of available devices is fixed, the modeler must increase $n_{\mathrm{pd}}$ (the number of model realizations on each device) to ensure there are sufficient Monte Carlo samples for reliable estimations. A naive implementation of ensemble simulation is simply running individual model realization one after another, whose time consumption increases linearly with the ensemble size. In this paper, we optimize the model efficiency by sharing cell parameters among different model realizations as shown in Fig. 3. We test the computation efficiency for different values of $n_{\mathrm{pd}}$. For each case, we use all 8 GPUs to run an ensemble of $8n_{\text{pd}}$ models, simulating a 1-hour inundation process. Meanwhile, the thread block dimension is optimized for each $n_{\text{pd}}$ value. As introduced in Fig. 1 (a 2-dimensional example), thread block is the basic program execution unit in CUDA. It has up to 3 dimensions ($dim_x, dim_y, dim_z$), and its size $dim_x dim_y dim_z$ should be limited to 256 as suggested by Nvidia. Here we set $dim_y = 1$ and $dim_z = n_{\text{pd}}$, and test different values for $dim_x$ for the best performance. Under these settings, the ensemble run time is measured and shown in Fig. 12 (a). When multiple models are run on the same device, the parallel paradigm in Fig. 3 helps reuse definitive cell parameters and reduce memory access overhead, thereby improving calculation efficiency and shortening the average time consumption of each ensemble member. We define $\bar{\eta}$ as the average run time of one ensemble member using one device:

$$\bar{\eta} = \frac{\eta_{\text{total}}}{n_{\text{pd}}} \tag{13}$$

where $\eta_{\text{total}}$ is the total time consumption of ensemble simulation, i.e., the y-axis of Fig. 12 (a). As demonstrated in Fig. 12 (b), the minimum $\bar{\eta}$ is achieved when $n_{\text{pd}} = 8$, with a 33% improvement compared with $n_{\text{pd}} = 1$. It is also discovered that $n_{\text{pd}}$ values of 6 and 10 result in inferior performances compared with values of 4, 8, and 12. This may imply that setting the block dimension to a multiple of 4 benefits thread parallelism.

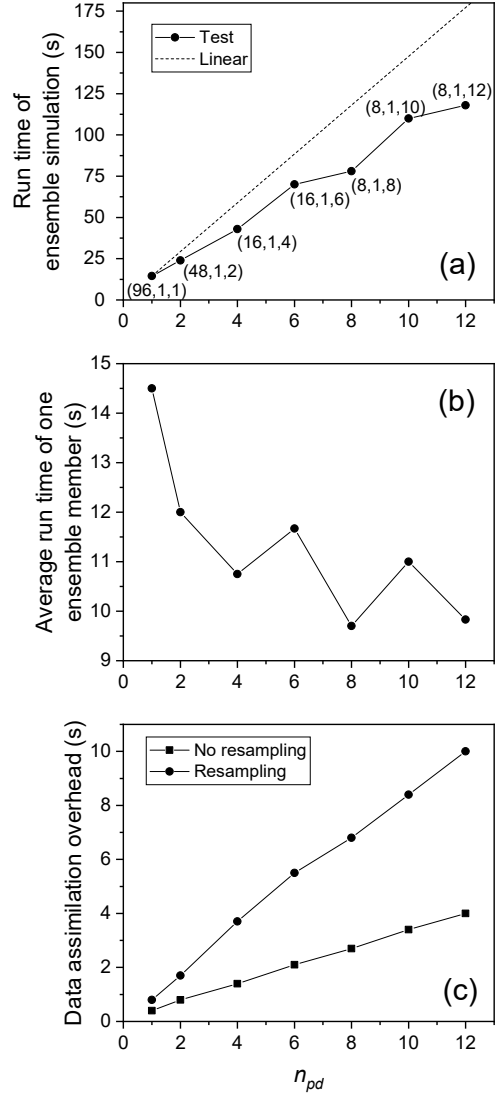

**Figure 12.** Ensemble-member scalability of the ensemble simulation and DA computation using 8 GPUs. The 3-tuples in subplot (a) indicate the optimum block dimension $(dim_x, dim_y, dim_z)$ for each value of $n_{\mathrm{pd}}$.

Besides, DA is operated based on synthetic observation data which is produced based on deterministic results. Three water depth gauges and two streamflow gauges are placed in the simulation region as illustrated in Fig. 6. The streamflow is calculated by accumulating the fluxes on the links across the observed section. Both the flow and water depth data are recorded every hour. Gaussian white noise with a 5% relative standard deviation is added to each data point as an artificial perturbation, generating the synthetic observation data. These data are assimilated into the model ensemble with a 1-hour interval using the Bayesian update scheme described in 2.2. We calculate the likelihood function in Eq. (9) using an error covariance matrix $\mathbf{\Sigma} = \text{diag}((\alpha d_1)^2, \ldots, (\alpha d_5)^2)$, with $d_1, \ldots, d_5$ being the current value of observations and $\alpha = 0.05$. After resampling, the ensemble is perturbed by multiplying the precipitation data of each particle with a factor $\beta \sim N(1.0, 0.05^2)$ to ensure particle diversity.

In Fig. 12 (c), we measure the overhead caused by a single round of assimilation calculations (Fig. 5) when different ensemble members and resampling strategies are involved. As the resampling process requires additional data transfer, it is also more time-demanding. Fig. 12 (c) indicates particle resampling increases the DA overhead by about 1.5 times. Meanwhile, it can also be seen that the DA time consumption grows almost linearly with the number of GPUs, which is expected considering the scale of data transferred.

To test the performance of the ensemble DA model, we use the assimilation strategy stated in 3.3 and assimilate the synthetic observations at the end of every hour. The 24-hour simulation and DA process for 96 members take 49.4 minutes using 8 GPUs. Compared with the simulation-only run, the assimilation of observations uses an additional 2.2 minutes and increases the total time consumption by 4.7%. The prior distribution of output and input factors after each time of assimilation are shown in Fig. 13 and Fig. 14 respectively. Significant ensemble variance can be observed in i.e., water depth predictions of gauge 2 at hour 8. Such diversities reduce with time and the simulated time series of inundation and streamflow keep close to the observation values, which implies that the assimilation algorithm successfully controls the model uncertainties and errors.

Meanwhile, the estimations derived for the input factors show higher levels of deviation from the reference value. The precipitation, as the most essential driving force, is constantly constrained in the neighborhood of the input series used in the deterministic case. The friction of water body bed is also estimated with satisfying accuracy, converging to the reference value after hour 12. On the other hand, the estimated roughness coefficients of land cells exhibit certain biases. Both the friction coefficients of green land and hardened ground have relative errors of 5% ~ 10%. Besides, the drainage capacity of pipe networks does not converge well with a stable ensemble spread throughout the simulation process. The inferior estimation results are mainly due to the equifinality of those parameters, whose effects on the model outputs are covered by more influential input factors such as rainfall and are less identifiable in assimilation. In general, the presented model identifies the major sources of uncertainty and effectively controls the accompanied simulation error.

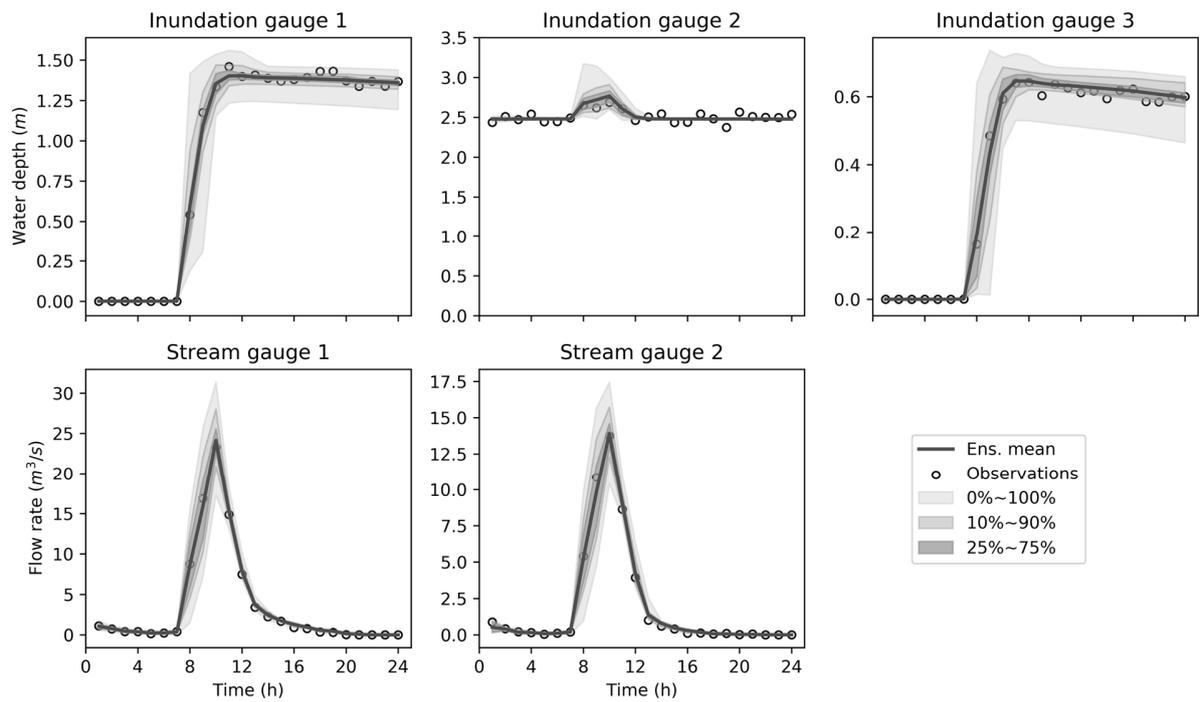

**Figure 13.** Probabilistic prediction results at different gauges. Shades with different colors represent different intervals of percentile.

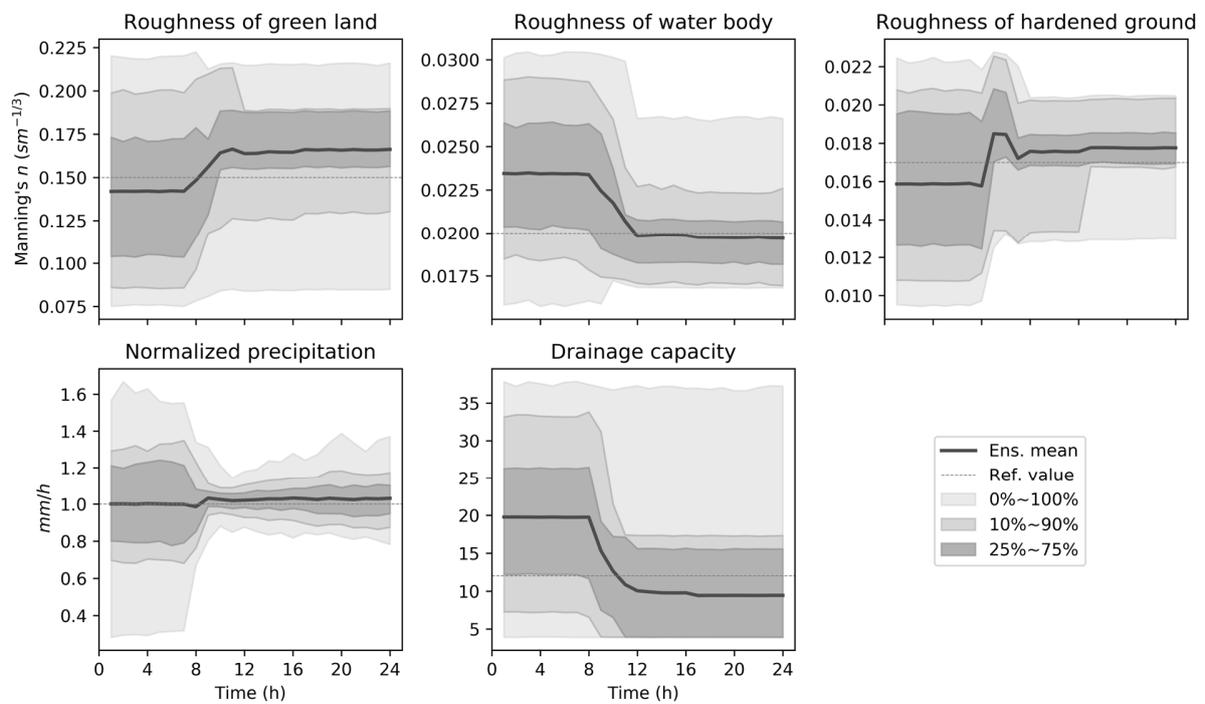

**Figure 14.** Temporal evolution of the probability distributions of input factors. Shades with different colors represent different intervals of percentile. The reference values indicate the inputs used in the deterministic run which generates the synthetic observation data. The normalized precipitation is calculated by dividing the precipitation input of a certain particle using the black line data in Fig. 7.

The abundant information provided by ensemble-based modeling enables a better insight into the inundation process and associated risks. For instance, the spatial distribution of the probability of a certain level of inundation can be directly illustrated as in Fig. 15. The regions with large areas of heavy inundation such as Region 1 and Region 2 should receive more attention and take corresponding strategies, while the orange-colored districts can be less concerned about being flooded. In addition, the high-resolution model supports elaborate analysis of local inundation status. The PDF of water depth in each cell can be derived from ensemble results, allowing us to evaluate the inundation status at different percentiles. Taking the two regions in Fig. 15 for example, we can further construct refined 3-D inundation maps of different risk levels as shown in Fig. 16. The 10th and 90th percentiles of maximum water depth distribution are demonstrated, representing optimistic and pessimistic situations respectively. This information is especially useful for local authorities and residents to take specific measures for each building and avoid unnecessary losses.

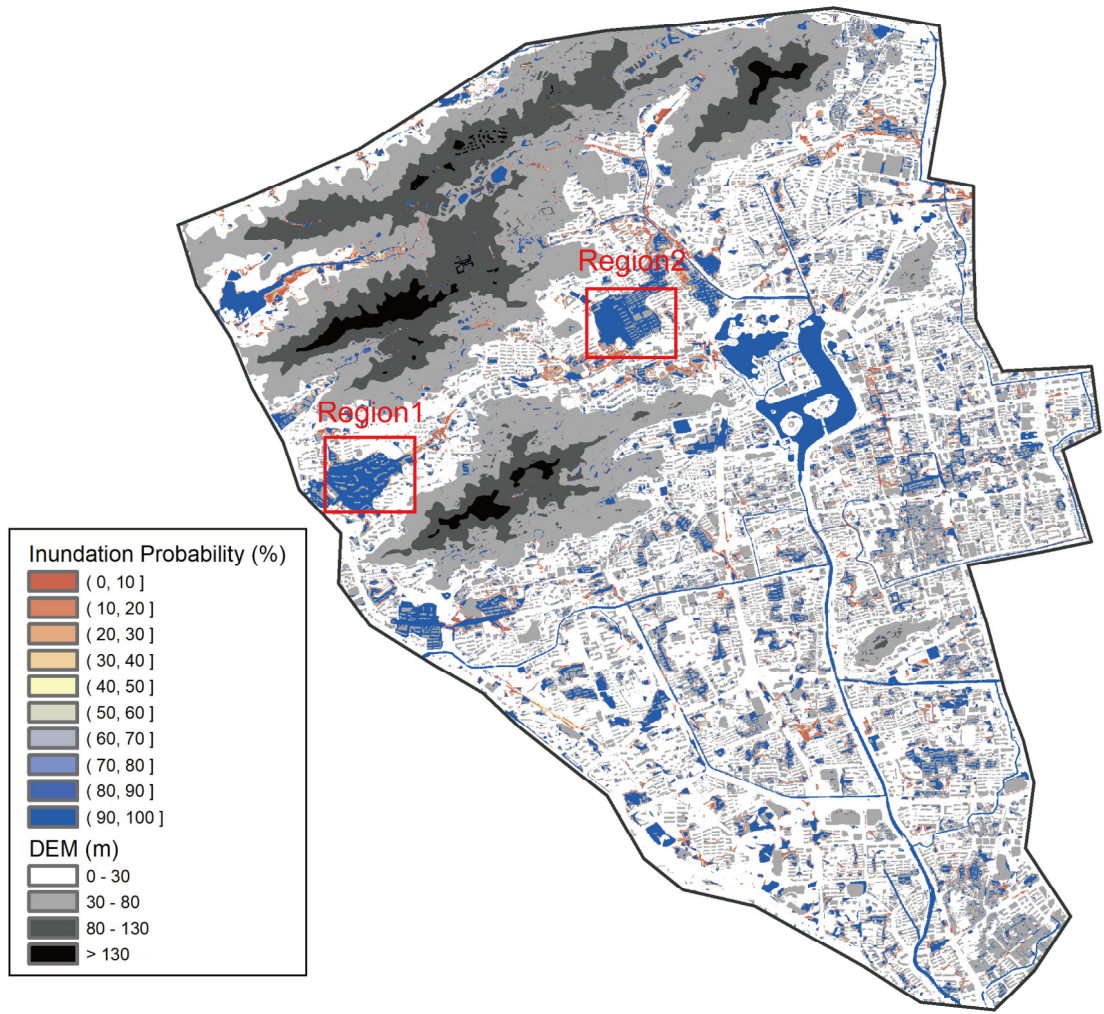

**Figure 15.** Spatial distribution of heavy-inundation probabilities derived from the ensemble simulation. The threshold of heavy inundation is defined as the maximum water depth reaching 20 cm.

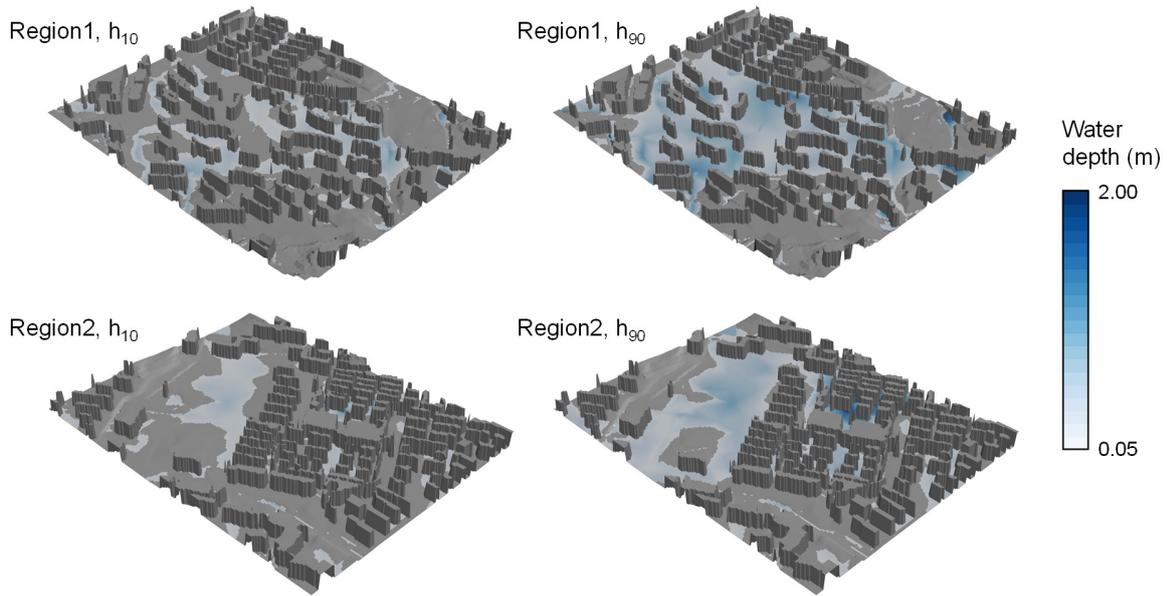

**Figure 16.** Local 3-D inundation maps of different risk levels. $h_a$ indicates the $a$th percentile of the water depth PDF.

## 5 Conclusions

In this paper, a real-time probabilistic flood inundation model based on high-performance computing techniques is developed. Although previous studies have investigated different approaches of parallelism (Sanders and Schubert, 2019; Xia et al., 2019) and data assimilation (Zarekarizi, 2018) in flood modeling, here we provide a detailed investigation on the implementation and optimization of a flood prediction and ensemble DA system on multi-GPU architectures, which have rarely been discussed. An ensemble of model realizations is constructed to represent the uncertainties involved in model inputs. We use multi-GPU parallelization and various optimization techniques to enhance computation efficiency for more than three orders of magnitude compared with the single-thread program on CPU. Meanwhile, the particle filtering algorithm is employed to assimilate observation data sequentially. Synthetic experiment results show that the adopted DA scheme can effectively control the model uncertainties and improve prediction accuracy. Future works may focus on integrating the presented approaches with numerical weather predictions and high-precision remote sensing data (e.g., Ming et al., 2020) to enhance the efficiency and reliability of operational flood forecast systems.

## Acknowledgements

This paper is supported by the National Key Research and Development Program of China (2018YFE0196000) and the National Natural Science Foundation of China (U2240203). The topographical and land cover data used in this paper can be found at https://osf.io/wng2f/. The authors declare no conflict of interest.

## References

Abbaszadeh, P., Gavahi, K., & Moradkhani, H. (2020). Multivariate remotely sensed and in-situ data assimilation for enhancing community WRF-Hydro model forecasting. Advances in Water Resources, 145, 103721.

Afzal, A., Ansari, Z., Faizabadi, A. R., & Ramis, M. K. (2017). Parallelization strategies for computational fluid dynamics software: state of the art review. Archives of Computational Methods in Engineering, 24(2), 337-363.

Alawadhi, A., Boso, F., & Tartakovsky, D. M. (2018). Method of distributions for water hammer equations with uncertain parameters. Water Resources Research, 54(11), 9398-9411.

Anderson, J. L., & Collins, N. (2007). Scalable implementations of ensemble filter algorithms for data assimilation. Journal of Atmospheric and Oceanic Technology, 24(8), 1452-1463.

Arulampalam, M. S., Maskell, S., Gordon, N., & Clapp, T. (2002). A tutorial on particle filters for online nonlinear/non-Gaussian Bayesian tracking. IEEE Transactions on signal processing, 50(2), 174-188.

Aureli, F., Prost, F., Vacondio, R., Dazzi, S., & Ferrari, A. (2020). A GPU-accelerated shallow-water scheme for surface runoff simulations. Water, 12(3), 637.

Barthélémy, S., Ricci, S., Rochoux, M. C., Le Pape, E., & Thual, O. (2017). Ensemble-based data assimilation for operational flood forecasting–On the merits of state estimation for 1D hydrodynamic forecasting through the example of the "Adour Maritime" river. Journal of Hydrology, 552, 210-224.

Bates, P. D., & De Roo, A. P. J. (2000). A simple raster-based model for flood inundation simulation. Journal of hydrology, 236(1-2), 54-77.

Bates, P. D., Horritt, M. S., & Fewtrell, T. J. (2010). A simple inertial formulation of the shallow water equations for efficient two-dimensional flood inundation modelling. Journal of Hydrology, 387(1-2), 33-45.

Cao, Y., Ye, Y., Liang, L., Zhao, H., Jiang, Y., Wang, H., ... & Yan, D. (2019). A modified particle filter-based data assimilation method for a high-precision 2-D hydrodynamic model considering spatial-temporal variability of roughness: Simulation of dam-break flood inundation. Water Resources Research, 55(7), 6049-6068.

Chaney, N. W., Herman, J. D., Reed, P. M., & Wood, E. F. (2015). Flood and drought hydrologic monitoring: the role of model parameter uncertainty. Hydrology and Earth System Sciences, 19(7), 3239-3251.

Chen, H., Yang, D., Hong, Y., Gourley, J. J., & Zhang, Y. (2013). Hydrological data assimilation with the Ensemble Square-Root-Filter: Use of streamflow observations to update model states for real-time flash flood forecasting. Advances in Water Resources, 59, 209-220.

Chen, J., Hill, A. A., & Urbano, L. D. (2009). A GIS-based model for urban flood inundation. Journal of Hydrology, 373(1-2), 184-192.

Cloke, H. L., & Pappenberger, F. (2009). Ensemble flood forecasting: A review. Journal of hydrology, 375(3-4), 613-626.

Cooper, E. S., Dance, S. L., García-Pintado, J., Nichols, N. K., & Smith, P. J. (2019). Observation operators for assimilation of satellite observations in fluvial inundation forecasting. Hydrology and Earth System Sciences, 23(6), 2541-2559.

Cozzolino, L., Cimorelli, L., Della Morte, R., Pugliano, G., Piscopo, V., & Pianese, D. (2019). Flood propagation modeling with the Local Inertia Approximation: Theoretical and numerical analysis of its physical limitations. Advances in Water Resources, 133, 103422.

Cunge, J. (1980). Practical aspects of computational river hydraulics. Pitman Publishing Ltd. London,(17 CUN), 1980, 420.

De Almeida, G. A., & Bates, P. (2013). Applicability of the local inertial approximation of the shallow water equations to flood modeling. Water Resources Research, 49(8), 4833-4844.

Dottori, F., & Todini, E. (2011). Developments of a flood inundation model based on the cellular automata approach: testing different methods to improve model performance. Physics and Chemistry of the Earth, Parts A/B/C, 36(7-8), 266-280.

Doucet, A., Godsill, S., & Andrieu, C. (2000). On sequential Monte Carlo sampling methods for Bayesian filtering. Statistics and computing, 10(3), 197-208.

Evensen, G. (2009). Data assimilation: the ensemble Kalman filter. Springer Science & Business Media.

Fan, Y., Ao, T., Yu, H., Huang, G., & Li, X. (2017). A coupled 1D-2D hydrodynamic model for urban flood inundation. Advances in Meteorology, 2017.

Gironás, J., Roesner, L. A., Rossman, L. A., & Davis, J. (2010). A new applications manual for the Storm Water Management Model(SWMM). Environmental Modelling & Software, 25(6), 813-814.

Govett, M., Rosinski, J., Middlecoff, J., Henderson, T., Lee, J., MacDonald, A., ... & Duarte, A. (2017). Parallelization and Performance of the NIM Weather Model on CPU, GPU, and MIC Processors. Bulletin of the American Meteorological Society, 98(10), 2201-2213.

Han, S., & Coulibaly, P. (2017). Bayesian flood forecasting methods: A review. Journal of Hydrology, 551, 340-351.

Hostache, R., Chini, M., Giustarini, L., Neal, J., Kavetski, D., Wood, M., ... & Matgen, P. (2018). Near-real-time assimilation of SAR-derived flood maps for improving flood forecasts. Water Resources Research, 54(8), 5516-5535.

Kalyanapu, A. J., Shankar, S., Pardyjak, E. R., Judi, D. R., & Burian, S. J. (2011). Assessment of GPU computational enhancement to a 2D flood model. Environmental Modelling & Software, 26(8), 1009-1016.

Kobayashi, K., Kitamura, D., Ando, K., & Ohi, N. (2015). Parallel computing for high-resolution/large-scale flood simulation using the K supercomputer. Hydrological Research Letters, 9(4), 61-68.

Kurtz, W., Lapin, A., Schilling, O. S., Tang, Q., Schiller, E., Braun, T., ... & Brunner, P. (2017). Integrating hydrological modelling, data assimilation and cloud computing for real-time management of water resources. Environmental modelling & software, 93, 418-435.

Kusch, J., & Frank, M. (2018). Intrusive methods in uncertainty quantification and their connection to kinetic theory. International Journal of Advances in Engineering Sciences and Applied Mathematics, 10(1), 54-69.

Lacasta, A., Morales-Hernández, M., Murillo, J., & García-Navarro, P. (2015). GPU implementation of the 2D shallow water equations for the simulation of rainfall/runoff events. Environmental Earth Sciences, 74(11), 7295-7305.

Liu, F., Wang, L., Li, X., & Huang, C. (2020). ComDA: a common software for nonlinear and non-Gaussian land data assimilation. Environmental Modelling & Software, 127, 104638.

Liu, Y., Weerts, A. H., Clark, M., Hendricks Franssen, H. J., Kumar, S., Moradkhani, H., ... & Restrepo, P. (2012). Advancing data assimilation in operational hydrologic forecasting: progresses, challenges, and emerging opportunities. Hydrology and Earth System Sciences, 16(10), 3863-3887.

Ming, X., Liang, Q., Xia, X., Li, D., & Fowler, H. J. (2020). Real-time flood forecasting based on a high-performance 2-D hydrodynamic model and numerical weather predictions. Water Resources Research, 56(7), e2019WR025583.

Moradkhani, H., Hsu, K. L., Gupta, H., & Sorooshian, S. (2005). Uncertainty assessment of hydrologic model states and parameters: Sequential data assimilation using the particle filter. Water resources research, 41(5).

Morales-Hernández, M., Sharif, M. B., Kalyanapu, A., Ghafoor, S. K., Dullo, T. T., Gangrade, S., ... & Evans, K. J. (2021). TRITON: A Multi-GPU open source 2D hydrodynamic flood model. Environmental Modelling & Software, 141, 105034.

Nerger, L., & Hiller, W. (2013). Software for ensemble-based data assimilation systems—Implementation strategies and scalability. Computers & Geosciences, 55, 110-118.

Nerger, L., Hiller, W., & Schröter, J. (2005). PDAF-the parallel data assimilation framework: experiences with Kalman filtering. In Use of high performance computing in meteorology (pp. 63-83).

Noh, S. J., Lee, S., An, H., Kawaike, K., & Nakagawa, H. (2016). Ensemble urban flood simulation in comparison with laboratory-scale experiments: Impact of interaction models for manhole, sewer pipe, and surface flow. Advances in Water Resources, 97, 25-37.

Pham, D. T. (2001). Stochastic methods for sequential data assimilation in strongly nonlinear systems. Monthly weather review, 129(5), 1194-1207.

Sanders, B. F., & Schubert, J. E. (2019). PRIMo: Parallel raster inundation model. Advances in Water Resources, 126, 79-95.

Sanders, B. F., Schubert, J. E., & Detwiler, R. L. (2010). ParBreZo: A parallel, unstructured grid, Godunov-type, shallow-water code for high-resolution flood inundation modeling at the regional scale. Advances in Water Resources, 33(12), 1456-1467.

Sharif, M. B., Ghafoor, S. K., Hines, T. M., Morales-Hernändez, M., Evans, K. J., Kao, S. C., ... & Gangrade, S. (2020, June). Performance Evaluation of a Two-Dimensional Flood Model on Heterogeneous High-Performance Computing Architectures. In Proceedings of the Platform for Advanced Scientific Computing Conference (pp. 1-9).

Shaw, J., Kesserwani, G., Neal, J., Bates, P., & Sharifian, M. K. (2021). LISFLOOD-FP 8.0: the new discontinuous Galerkin shallow-water solver for multi-core CPUs and GPUs. Geoscientific Model Development, 14(6), 3577-3602.

Shustikova, I., Domeneghetti, A., Neal, J. C., Bates, P., & Castellarin, A. (2019). Comparing 2D capabilities of HEC-RAS and LISFLOOD-FP on complex topography. Hydrological Sciences Journal, 64(14), 1769-1782.

Smith, L. S., & Liang, Q. (2013). Towards a generalised GPU/CPU shallow-flow modelling tool. Computers & Fluids, 88, 334-343.Sridharan, B., Gurivindapalli, D., Kuiry, S. N., Mali, V. K., Nithila Devi, N., Bates, P. D., & Sen, D. (2020). Explicit expression of weighting factor for improved estimation of numerical flux in Local Inertial models. Water Resources Research, 56(7), e2020WR027357.

Teng, J., Jakeman, A. J., Vaze, J., Croke, B. F., Dutta, D., & Kim, S. (2017). Flood inundation modelling: A review of methods, recent advances and uncertainty analysis. Environmental modelling & software, 90, 201-216.

Vacondio, R., Dal Palù, A., & Mignosa, P. (2014). GPU-enhanced finite volume shallow water solver for fast flood simulations. Environmental modelling & software, 57, 60-75.

Vivoni, E. R., Mascaro, G., Mniszewski, S., Fasel, P., Springer, E. P., Ivanov, V. Y., & Bras, R. L. (2011). Real-world hydrologic assessment of a fully-distributed hydrological model in a parallel computing environment. Journal of Hydrology, 409(1-2), 483-496.

Wang, Y., Chen, A. S., Fu, G., Djordjević, S., Zhang, C., & Savić, D. A. (2018). An integrated framework for high-resolution urban flood modelling considering multiple information sources and urban features. Environmental modelling & software, 107, 85-95.

Whitaker, J. S., & Hamill, T. M. (2002). Ensemble data assimilation without perturbed observations. Monthly weather review, 130(7), 1913-1924.

Wu, W., Emerton, R., Duan, Q., Wood, A. W., Wetterhall, F., & Robertson, D. E. (2020). Ensemble flood forecasting: Current status and future opportunities. Wiley Interdisciplinary Reviews: Water, 7(3), e1432.

Xia, X., Liang, Q., & Ming, X. (2019). A full-scale fluvial flood modelling framework based on a high-performance integrated hydrodynamic modelling system (HiPIMS). Advances in Water Resources, 132, 103392.

Xu, X., Zhang, X., Fang, H., Lai, R., Zhang, Y., Huang, L., & Liu, X. (2017). A real-time probabilistic channel flood-forecasting model based on the Bayesian particle filter approach. Environmental Modelling & Software, 88, 151-167.

Zarekarizi, M. (2018). Ensemble Data Assimilation for Flood Forecasting in Operational Settings: From Noah-MP to WRF-Hydro and the National Water Model (Doctoral dissertation, Portland State University).

Zhou, Y., Shen, D., Huang, N., Guo, Y., Zhang, T., & Zhang, Y. (2019). Urban flood risk assessment using storm characteristic parameters sensitive to catchment-specific drainage system. Science of the Total Environment, 659, 1362-1369.

Ziliani, M. G., Ghostine, R., Ait-El-Fquih, B., McCabe, M. F., & Hoteit, I. (2019). Enhanced flood forecasting through ensemble data assimilation and joint state-parameter estimation. Journal of Hydrology, 577, 123924.